\documentclass[12pt]{article}

\usepackage[tbtags]{amsmath}
\usepackage{amssymb}
\usepackage{amsfonts}
\usepackage{euscript}
\usepackage{longtable}
\usepackage{graphicx}
\usepackage{color}
\usepackage{shadow}

\begin{document}

\title{\sc Constructing  $\mathrm{SU(2)} \times  \mathrm{U(1)}$ orbit space
for qutrit mixed states}

\author{Vladimir  Gerdt\, ${}^{a}$\,,  Arsen  Khvedelidze\,${}^{a, b}$\,  and  Yuri  Palii \,${}^{a,c}$
\\[1cm]
${}^a$\,\small\it Laboratory of  Information Technologies,  Joint Institute for Nuclear Research, Dubna, Russia\\
${}^b$\,\small\it Iv. Javakhishvili Tbilisi State University,  A.Razmadze Mathematical Institute,  Georgia \\
${}^c$\,\small\it Institute of Applied Physics, Moldova Academy of Sciences,
Chisinau, Republic of Moldova
}

\date{\empty }

\maketitle

\begin{abstract}
 The orbit space $\mathfrak{P}(\mathbb{R}^8)/\mathrm{G} \,$  of the group $\mathrm{G}:=\mathrm{SU(2)\times U(1)}  \subset \mathrm{U(3)}$  acting adjointly  on the state space $\mathfrak{P}(\mathbb{R}^8)$  of  a  3-level quantum system is discussed. The semi-algebraic structure of
$\mathfrak{P}(\mathbb{R}^8) /\mathrm{G} \,  $  is determined
within the  Procesi-Schwarz  method.  Using the integrity  basis for  the ring of  G\--invariant polynomials,  $\mathbb{R}[\mathfrak{P}(\mathbb{R}^8)]^{\mathrm{G}}\,, $ the set of constraints on the  Casimir invariants  of  $\mathrm{U}(3)$ group coming from the  positivity requirement of
Procesi-Schwarz gradient matrix,  $\mathrm{Grad}(z) \geqslant 0\,,$  is  analyzed in details.
\end{abstract}

\newpage

\tableofcontents

\newpage

\section{Introduction}

Since a  very beginning of quantum mechanics,
a  highly nontrivial interplay between the quantities describing a composite quantum system as a ``single whole''  and ``local characteristics''  of its constituents  became the subject of intensive  studies (holistic v.s. reductionism views).
The present note aims to discuss  a mathematical aspect of
``the whole and the parts''  problem in quantum theory
considering a model of 3-dimensional quantum system, qutrit.
Skipping aside the physical motivation,   these mathematical issues  can be formulated as follows.

Consider  the compact Lie group $\mathrm{G}$  acting on a real $n$\--dimensional  space $V$ and let
$\mathrm{H}\subset\mathrm{G} $ is  its compact subgroup.
Assume that  the corresponding orbit spaces $V/\mathrm{G}$ and $V/\mathrm{H}\,$  admit  a realization as
semi-algebraic subsets, $Z(V/\mathrm{G})$ and $Z(V/\mathrm{H})$ of $\mathbb{R}^{q}\,$  for a certain $q$.
The mathematical  version of  ``the whole and the parts''  dilemma   can be formulated as
the problem of determination of correspondence between sets   $Z(V/\mathrm{H})$  and  $Z(V/\mathrm{G})$.

In applications to the quantum theory the role of space $V$ plays  the space  of mixed states of $n$\--dimensional  binary quantum system, $\mathfrak{P}(\mathbb{R}^{n^2-1})$.
The groups $\mathrm{G}$ and $\mathrm{H}$ are  associated with the unitary  group  $\mathrm{U}(n)$ and its subgroup,   $\mathrm{U}(n_1)\times\mathrm{U}(n_2) \subset \mathrm{U}(n)\,, $
\footnote{The subgroup $H$ is determined by a fixed decomposition of  system
onto the $n_1$\-- and $n_2$\-- dimensional  subsystems,  such that $n=n_1\times n_2$.}
 acting in adjoint manner
 \begin{equation}\label{eq:Adjaction}
   \mathrm{ Ad}\left( g \right)\varrho = g \varrho g^{-1} \qquad g\in \mathrm{U}(n)
\end{equation}
 on the density matrices  $\varrho \in \mathfrak{P}(\mathbb{R}^{n^2-1}).$ The action (\ref{eq:Adjaction}) determines  the  `` global orbit space'',  $\mathfrak{P}(\mathbb{R}^{n^2-1})\,| \mathrm{U}(n)\,,$  and
the so-called \textit{entanglement space} $ \mathfrak{P}(\mathbb{R}^{n^2-1})\,|  \mathrm{U}(n_1)\times\mathrm{U}(n_2)$ of  a binary $n_1\times n_2$ system.

The semi-algebraic structure of  both orbit spaces admits description in terms of  the corresponding ring of
$\mathrm{G}$\--invariant polynomials, $\mathbb{R}[\mathfrak{P}]^{\mathrm{U}(n)}\,$ and $\mathbb{R}[\mathfrak{P}]^ {\mathrm{U}(n_1)\times\mathrm{U}(n_2)}\,$.
According to the   Procesi and Schwarz method \cite{ProcesiSchwarz1985PHYSLETT,ProcesiSchwarz1985}
these  semi-algebraic varieties in $\mathbb{R}^{q}\,$ are
defined  by the syzygy ideal  for the corresponding integrity basis and the semi-positivity of  the  so-called gradient matrix, $\mathrm{Grad}(z) \geqslant 0\,.$
 As it was discussed recently in \cite{GKP2014},  the orbit space   $\mathfrak{P}(\mathbb{R}^{n^2-1})\,| \mathrm{U}(n)\, $ representation in terms of the integrity basis for U(n)\--invariant polynomial ring  is completely  determined  from the physical requirements formulated as  the semi-positivity and Hermicity of
 density matrices. The conditions $\mathrm{Grad}(z) \geqslant 0\,$ do not bring any new restriction on
the elements in the  integrity basis for  $\mathbb{R}[\mathfrak{P}]^{\mathrm{U}(n)}\,.$
In contrast to that case,
the algebraic and geometric properties of the entanglement space, are more subtle.
It turns that in order to determine the local orbit space
$\mathfrak{P}(\mathbb{R}^{n^2-1})\,|  \mathrm{U}(n_1)\times\mathrm{U}(n_2)$ the additional constraints
arising from the semi-positivity of  $\mathrm{Grad}$\--matrix should be taken into account.
Moreover additional inequalities  in  elements of the integrity basis for $\mathbb{R}[\mathfrak{P}]^{\mathrm{U}(n_1)\times\mathrm{U}(n_2)}$ provide constraints on the U(n)\--invariants. Below, aiming to exemplify this statement the toy model, which mimicry  a  generic case of a binary composite system will be studied.  Namely,  we consider the 3-dimensional quantum system, defined by the state space $\mathfrak{P}(\mathbb{R}^{8}) \,,$  which  is a locus in quo of  the action of the symmetry group U(3) and  its U(2) subgroup $\mathrm{SU}(2)\times\mathrm{U}(1)\,.$

\section{Qutrit}

\noindent{$\bullet$ {\bf  The qutrit state  parametrization } $\bullet$}
Consider  a quantum 3-level system, named  the  qutrit.
Its state, the semi-positive  Hermitian, of trace one matrix  $\varrho$ can be parameterized as follows:
\begin{equation}\label{eq:qutri}
\varrho = \displaystyle\frac{1}{3}\, \left(\mathbb{I}_3+\sqrt{3}\,
\sum_{a=1}^{8}\xi_a \lambda_a
\right)\,.
\end{equation}
Here the real parameters $\{\xi_a\}_{a=1,\ldots,8} $  are components of
the  8-dimensional Bloch vector $\boldsymbol{\xi}\,$ and
$\{\lambda_a\}_{a=1,\ldots,8} $  are the
 Gell-Mann matrices  generating   the Hermitian basis of  the Lie algebra
$\mathfrak{su}(3)\,$:
\begin{equation*}\label{eq:lam-matr}
\begin{array}{c}
\begin{array}{ccc}
 \lambda _{1}=\left(\begin{array}{ccc}
   0 & 1 & 0 \\
   1 & 0 & 0 \\
   0 & 0 & 0 \
 \end{array}\right)&
 \lambda _{2}=\left(\begin{array}{ccc}
   0 & -i & 0 \\
   i & 0 & 0 \\
   0 & 0 & 0 \
 \end{array}\right)&
  \lambda _{3}=\left(\begin{array}{ccc}
   1 & 0 & 0 \\
   0 & -1 & 0 \\
   0 & 0 & 0 \
 \end{array}\right)\\ & & \\
  \lambda _{4}=\left(\begin{array}{ccc}
   0 & 0 & 1 \\
   0 & 0 & 0 \\
   1 & 0 & 0 \
 \end{array}\right)&
 \lambda _{5}=\left(\begin{array}{ccc}
   0 & 0 & -i \\
   0 & 0 & 0 \\
   i & 0 & 0 \
 \end{array}\right)&
  \lambda _{6}=\left(\begin{array}{ccc}
   0 & 0 & 0 \\
   0 & 0 & 1 \\
   0 & 1 & 0 \
 \end{array}\right)\
 \end{array}
 \\ \\
\begin{array}{cc}
\lambda _{7}=\left(\begin{array}{ccc}
   0 & 0 & 0 \\
   0 & 0 & -i \\
   0 & i & 0 \
 \end{array}\right)&
  \lambda _{8}=\displaystyle{\frac{1}{\sqrt{3}}}\left(
  \begin{array}{ccc}
   1 & 0 & 0 \\
   0 & 1 & 0 \\
   0 & 0 & -2 \
\end{array}\right)
\end{array}\
\end{array}
\end{equation*}
The product of  two  Gell-Man matrices involves two basic sets of  $\mathfrak{su}(3)$ algebra constants:
\begin{equation}\label{eq:pairsLambda}
    \lambda_a\lambda_b=\displaystyle \frac{2}{3}\delta_{ab}+ \left( d_{abc}+ i f _{abc}\right)\lambda_c\,,
\end{equation}
where $ d_{abc}$ and $ f_{abc}$ denote components of the completely symmetric and skew-symmetric symbols defined  via the anti-commutators
$\{ , \}$  and commutators $[ , ]$ of the Gell-Mann matrices:
\[
   d_{abc}
   = \displaystyle\tfrac{1}{4}\mbox{Tr}(\{\lambda_a,\lambda_b\}\lambda_c)\,, \qquad
 f_{abc}
   = \displaystyle\tfrac{1}{4}\mbox{Tr}([\lambda_a,\lambda_b]\lambda_c)  \,.
\]

The matrix $\varrho$ from (\ref{eq:qutri}) represents a physical state of qutrit iff
the Bloch vector $\boldsymbol{\xi}$ is subject to the  following polynomial  constraints:
\footnote{The inequalities   (\ref{eq:pos1}) and   (\ref{eq:pos2}) reflect  the semi-positivity of
 qutrit's density matrices,  $\varrho \geq 0\,.$
}
\begin{eqnarray}
\label{eq:pos1}
  &&\xi_a\xi_a \leq 1\,, \\
   \label{eq:pos2}
  && 0\leq  \xi_a\xi_a -\displaystyle\frac{2}{\sqrt{3}} d_{abc}\xi_a\xi_b\xi_c\, \leq \frac{1}{3}\,,
\end{eqnarray}

\noindent{$\bullet$ {\bf  The unitary symmetry of qutrit}
$\bullet$} As it was mentioned above the  unitary group U(3) acts on  $\mathfrak{P}(\mathbb{R}^8)$ in adjoint manner.
The Bloch vector $\boldsymbol{\xi}$  transforms  under Ad\---action as
8-dimensional vector
\[
\xi_a^\prime= O_{ab}\xi_b \,, \qquad\qquad O \in SO(8)\,,
\]
with the special 8-parametric subgroup of $SO(8)\,.$
\footnote{More details on the algebraic  and geometric structures of  the  SU(3) group can be found in
 the classical paper \cite{MichelRadicati}.}

\noindent{$\bullet$ {\bf  The ``local symmetry''  $\mathrm{SU}(2)\times \mathrm{U}(1) $} $\bullet$}
Consider  the U(2) subgroup of U(3) identified (up to conjugation) by
the conventional embedding:
\begin{equation}\label{eq:U2xU1}
   U(2)=\left\{g(u) =\left(\begin{array}{c|c}
            u & \begin{array}{c}\vspace{0.2cm} \\
            \vspace{0.2cm} \end{array}
             \\ \hline
            \begin{array}{cc} \hspace{0.4cm} & \vspace{0.4cm} \end{array}
             & (\det u)^{-1} \\
          \end{array}\right)\  |  \  u\in U(2)  \right\} \subset SU(3)\,.
\end{equation}
According to the embedding (\ref{eq:U2xU1}) and to the Gell-Mann basis choice,
the U(2) subgroup   is generated by $\lambda_1, \lambda_2, \lambda_3 $
(generators of SU(2) subgroup) and $\lambda_8$ (generator of U(1) subgroup).  An element of U(2) subgroup can be written as
\begin{equation}\label{eq:subgroupParame}
    g = \exp(i\lambda_1\alpha)\exp(i\lambda_2\beta)\exp(i \lambda_3\gamma)\exp (i\theta\lambda_8)\,,
\end{equation}
where the Euler angles $\alpha, \beta, \gamma $ parametrize the SU(2) group and angle $\theta$  corresponds  to the U(1) subgroup phase,
 $\det u = \exp(i\frac{2}{\sqrt 3} \theta)\,.$

\section{Sketch of the Procesi-Schwarz method }
\label{sec:Proc-Schwarz}

The Classical theory of Invariants represents the cornerstone  in  description of orbit spaces.
Based on this theory
(see .e.g.  \cite{ProcesiBook}) the basic ingredients of
the description can be formulated as follows.

Consider  the compact Lie group G acting  linearly  on the real $d$\--dimensional vector space $V$.
 Let $\mathbb{R}[V]^{\mathrm{G}}$ is the corresponding ring of the $\mathrm{G}$\--invariant polynomials
on $V$.
Assume  ${\cal P} = \left(p_1, p_2, \dots ,p_q\right)$  is a  set of homogeneous polynomials
that form the  integrity basis,
$$\mathbb{R}[x_1, x_2,\dots, x_d]^{\mathrm{G}}= \mathbb{R}[p_1, p_2, \dots , p_q]\,.$$
Elements of the integrity basis define the polynomial mapping:
  \begin{equation}
  \label{eq:polmap}
p:  \qquad V \rightarrow\mathbb{R}^q\, ; \qquad  (x_1, x_2, \dots , x_d)  \rightarrow (p_1, p_2, \dots , p_q)\,.
\end{equation}
Since  $p$ is constant on the orbits of $\mathrm{G}$, it
 induces a homeomorphism of the orbit space $V/G$  and the image $X$ of  $p$\--mapping;  $V/G\simeq X$ \cite{CoxLittleO'Shea}.
In order to describe $X$  in terms of ${\cal P}$  uniquely,  it is necessary to take into  account  the
\textit{syzygy ideal}:
$$
 I_{\cal P}=\{h \in \mathbb{R}[   y_1, y_2, \dots , y_q]:  h(p_1, p_2, \dots, p_q) =0\,, \  \mathrm{in } \  \mathbb{R}[V ]\, \}.
$$
Let  $Z \subseteq  \mathbb{R}^q$  denote the  locus of common zeros of all elements of  $I_{\cal P}$.  Then $Z$ is algebraic subset of $\mathbb{R}^q\,$ such that $X \subseteq Z\,.$ Denoting by  $\mathbb{R}[ Z]$ the restriction of
 $\mathbb{R}[   y_1, y_2, \dots , y_q]$ to $Z$   one can easily verify that  $\mathbb{R}[ Z ]$ is
isomorphic to  the quotient
\(
\mathbb{R}[   y_1, y_2, \dots , y_q]/I_{\cal{P}} \) and thus $\mathbb{R}[Z] \simeq \mathbb{R}[V]^{\mathrm{G}}\,.$
Therefore the subset  $Z$  essentially is determined by $\mathbb{R}[V]^{\mathrm{G}}$,
but to describe  $X $ the further steps are required.
According to  \cite{ProcesiSchwarz1985PHYSLETT,ProcesiSchwarz1985} the necessary information on $X$ is encoded in the
structure of $q\times q $ matrix  with  elements  given by  the  inner products of
gradients, $\mathrm{grad}( p_i ):$
\begin{equation}
\label{eq:Grad}
||\mathrm{Grad}||_{ij}= \left(\mathrm{grad}\left(  p_i \right), \mathrm{grad}\left(  p_j \right)\right)\,.
\end{equation}

Summarizing all above  observations, the orbit space  can be  identified with
the semi-algebraic variety, defined as points, satisfying two conditions:
\begin{itemize}
\item[a)]
$z\in Z $,  where $Z$ is the surface defined by the syzygy ideal  for the integrity basis
in  $\mathbb{R}[V]^{\mathrm{G}}$;
\item[b)] $\mathrm{Grad}(z) \geqslant 0\,.$
\end{itemize}

\section{Constructing the G\--invariant polynomials  }

Let  $\mathrm{ GL}(n, \mathbb{C})\,$ be  the
general linear group of degree $n$ over the field $\mathbb{C}\,.$  Assume that $\mathrm{ GL}(n, \mathbb{C})\,,$ operates with  the polynomials $p(x_1, x_2, \dots , x_n) \in
\mathbb{C}[x_1, x_2, \dots , x_n]$   as follows:
\begin{equation}\label{eq:linactpol}
 (g p) \left(x_1, x_2, \dots , x_n\right)  :=  p\left(x^\prime_1, x^\prime_2, \dots , x^\prime_n\right)\,, \qquad
     g  \in \mathrm{ GL}(n, \mathbb{C})\,,
\end{equation}
where
\begin{equation}\label{eq:GLtr}
    x^\prime_i= g^{-1}_{ij}x_j\,.
\end{equation}
The polynomials  $p(x_1, x_2, \dots , x_n) $  are  called  G\--invariant if they represent  the fixed points of
 transformations  (\ref{eq:linactpol}):
\begin{equation}\label{eq:invpol}
    (g p)\left(x_1, x_2, \dots , x_n\right)  :=  p\left(x_1, x_2, \dots , x_n\right)\,.
\end{equation}
Here we are concerned with the polynomials in $n^2$  complex entries  of the density matrices  $p(\varrho) = p(\varrho_{11} , \varrho_{12}, \dots , \varrho_{nn} ) $.
To reduce the adjoint action
(\ref{eq:Adjaction}) to a linear transformation of the type  (\ref{eq:GLtr}) one can
identify the Hermitian density matrix $\varrho $ with
the complex vector  $V$ of length $n^2$  and consider the linear representation of the subgroup $L\subset GL(n, \mathrm{C})$ defined via  tensor product of unitary matrix  with its complex conjugated one
\begin{equation}\label{eq:Linearized}
   L :=
\mathrm{U(n)}\otimes\overline{\mathrm{U(n)}}\,.
\end{equation}
The invariant polynomials (\ref{eq:invpol}) form an algebra over the $\mathbb{C}$,
and any such invariant can be expressed as a polynomial of the so-called fundamental invariants,
the homogeneous polynomials of fixed degrees.
Since the  homogeneous invariants of a fixed degree form a vector space,  it is
sufficient to  find a  maximal, linearly independent set of homogeneous invariants,
 i.e., a basis for that vector space.
The dimension of  this vector space can be extracted from the power series (Poincare series \cite{PopovItogi}) expansion of the Molien function \cite{Forger}. In fact, given a compact Lie group $G$ and its representation $\pi$, the Molien function can be directly defined by the power series (cf.~\cite{Forger})
\begin{equation}
   M_{\pi}(\mathbb{C}[V]^{\mathrm{G}},q)=\sum_{k=0}^\infty c_k(\pi)q^k\,.   \label{PoincareSeries}
\end{equation}
Here $c_k(\pi)$ is the number of linearly independent
G-invariant polynomials of degree $k$ on $V$.

\subsection{The Molien function}

The Molien function (\ref{PoincareSeries}) associated to the representation $\pi(g)$  of a compact Lie group $G$ on $V\,$
 admits integral representation~\cite{Forger,ProcesiBook} (Molien's formula):
\begin{equation}\label{eq:Molienfunct}
    M_{\pi}(\mathbb{C}[V]^{\mathrm{G}},q)=\int_G\, \frac{d\mu(g)}{\det(\mathbb{I}-q\pi(g))} \qquad
|q|<1\,,
\end{equation}
where  ${d\mu(g)}$ is the Haar measure for Lie group $G\,.$  According   to  the Weyl's  Integration Formula \cite{ProcesiBook},
an integral over a compact Lie group G can be decomposed into a double integral over
a maximal torus T and over the quotient G/T of the group by its torus.
If the integrand is a  function invariant under conjugation in the group, then the
latter integral is ``q-independent'' and the total integral reduces to an integral  over the
maximal torus with coordinates $x$ and the additional Weyl factor $A(x)$:
\begin{equation}\label{eq:Molienfunct}
    M_{\pi}(\mathbb{C}[V]^{\mathrm{G}},q)=\int_T\, \frac{d\mu[x]\, A(x)}{\det(\mathbb{I}-q\pi(x))}\,,
\end{equation}
The resulting integral over the torus can be transformed into a complex
path integral and  evaluated using the residue theorem.

 In what follows we present the  Molien functions for  the U(3) and  its  U(2) subgroup on complex $9$-dimensional vectors accordingly to (\ref{eq:Linearized}).

\noindent{$\bullet${ \bf The Molien function for U(3)} $\bullet$ }
For the group U(3)  the Weyl factor $A(x)$ is squire of Vandermonde
determinant  calculated  for torus coordinates
divided by the order of the corresponding Weyl group:
\[
A_{{}_{\mathrm{SU(3)}}}(x_1, x_2,  x_3)= \frac{1}{3!}\,\prod_{i<j}^{3}(x_i-x_j)\overline{(x_i-x_j)}\,,
\]
and the Molien function is given by
\begin{equation}\label{eq:MolienSU3}
M^{(d=9)}_{ \mathrm{U}(3)}(q)=
\frac{1}{(1-q)(1-q^2)(1-q^3)}\,.
\end{equation}

\noindent{$\bullet$ {\bf  The Molien function for SU(2)$\times$U(1)} $\bullet$}
For  this case $\pi\otimes\bar{\pi}$ representation for maximal torus reads
\begin{equation*}
\begin{split}
\pi\otimes\bar{\pi}&=
(x,x^{-1},y)\otimes(x^{-1},x,y^{-1}) \\
& = (1, x^2, x y^{-1}, x^{-2}, 1, x^{-1} y^{-1}, y x^{-1}, x y, 1)\,,
\end{split}
\end{equation*}
where $x$ is coordinate on SU(2) group torus and $y$ is coordinate on U(1)\,.
The Weyl factor for SU(2) group
\[
A{{}_{\mathrm{SU(2)}}}(x):= 1- \frac{x^2-x^{-2}}{2}
\]
implies reduction of  (\ref{eq:Molienfunct})
to the double path integral
\[
\begin{split}
M^{(d=9)}_{ \mathrm{SU}(2)\times \mathrm{U}(1)}(q)&=
\int\frac{d\,\mu_{\mathrm{SU}(2)}d\,\mu_{\mathrm{U}(1)}}
{\det|1-q\,\pi\otimes\bar{\pi}|} \\
& =\frac{1}{8\pi^2}\frac{1}{(1-q)^3}
\oint_{|x|=1} \oint_{|y|=1}\frac{(1-x^2)^2\,xdx \, ydy}
{(1-q x^2)(1-q x y)(y-q x)(x-q y)(xy-q)(x^2-q)}
\end{split}.
\]
Subsequent calculation of the residues of the integrand, at  first with respect to $y$ at poles
\(P_y=\{qx,\,q/x\}
\)
and then with respect to $x$ variable at poles
\(P_x=\{\pm\sqrt{q},\, \pm q\},
\)
gives finally the rational expression for the
Molien function:
\begin{equation}\label{eq:Msu2u1}
M^{(d=9)}_{ \mathrm{SU}(2)\times \mathrm{U}(1)}(q)=
\frac{1}{(1-q)(1-q^2)^2(1-q^3)}\,.
\end{equation}

\subsection{U(3) and SU(2)$\times$U(1)\--invariant polynomials }

Expressions (\ref{eq:MolienSU3}) and (\ref{eq:Msu2u1}) for Molien functions indicate that the set fundamental homogeneous polynomials for rings $\mathbb{C}[x]^{\mathrm{SU(3)}}$
consists of three polynomials of degree 1, 2 and 3, while there  are five
$\mathrm{SU(2)\times U(1)}$\--invariant homogeneous  polynomials forming the
integrity basis  for he ring
 $\mathbb{C}[x]^{\mathrm{SU(2)\times U(1)}}\,.$  The latter basis includes one polynomial of degree 1, two polynomials of degree 2 and one polynomial of degree 3.

As the  integrity basis for the ring $\mathbb{C}[x]^{\mathrm{SU(3)}}$ can be composed either of the trace invariants
$t_k=\mathrm{tr}\left(\varrho^k\right)\,, k=1,2,3$ or of the SU(3) Casimir  invariants constructed   via correspondence  with the elements of the center
universal  enveloping algebra $\mathfrak{U}(\mathfrak{su}(3))$.

\noindent$\bullet$ {\bf Casimir invariants  }$\bullet$ Accordingly to the Bloch parametrization for the qutrit's density matrix
(\ref{eq:qutri}), the first order Casimir is fixed, $\mathrm{tr}\varrho =0$, while
the quadratic and qubic Casimir invariants are
the following polynomials
\begin{eqnarray}
  \mathfrak{C}_2 &=&\xi_i \xi_i\,, \\
   \mathfrak{C}_3 &=& \sqrt{3}\,d_{ijk}\xi_i \xi_j \xi_k \,,
\end{eqnarray}

\noindent$\bullet$ {\bf   $\mathrm{SU}(2)\times \mathrm{U}(1)$\--invariants   }$\bullet$
 The graded structure of the ring of invariants allows to construct its basis using  homogeneous polynomials of certain degrees.  These  homogeneous  G -invariant polynomials of a given degree  are defined as solution the system  of linear homogeneous equations (\ref{eq:invpol}).  Actually those equations are reduced to their infinitesimal version of the following form \cite{PopovItogi}
 \begin{eqnarray*}
&& e_i f =0\,,\quad i=1, \dots , m\,,\\
&& g_j f = f\,, \quad i=1, \dots , s \,,
\end{eqnarray*}
where $e_1, \dots,  e_m$ form the basis of Borel subgroup $B \subset G $ and $g_1, \dots,  g_s$ is a  system of representatives of conjugated classes for the group $G$ with respect to its connected subgroup $G^0\,.$
Applying this  generic  scheme  one can derive the following set of $\mathrm{SU}(2)\times \mathrm{U}(1)$\--invariants:
\begin{eqnarray}
\label{eq:f1}
 f_1 & = &\xi_8\,, \\
  f_2& = &\xi_1^2 + \xi_2^2 + \xi_3^2 \,,\\
   \label{eq:f2}
   f_3  & = &\xi_4^2 + \xi_5^2 + \xi_6^2 + \xi_7^2 \,,\\
     \label{eq:f3}
   f_4 & = &2(-\xi_1(\xi_4\xi_6+\xi_5\xi_7)+\xi_2(\xi_4\xi_7-\xi_5\xi_6)) +\xi_3(-\xi_4^2-\xi_5^2+\xi_6^2+\xi_7^2)\,.
   \label{eq:f4}
\end{eqnarray}

\section{Orbit spaces of qutrit}

Before  applying  the  above mentioned method by Processi and Schwarz \cite{ProcesiSchwarz1985PHYSLETT,ProcesiSchwarz1985}  to the orbit space construction
 let us reformulate  the  semi-algebraic description of the qutrit  state space
$\mathfrak{P}(\mathbb{R}^{8}) \,$ in terms of the
SU(3) Casimir invariants. In doing so, we mainly follow the ideology presented in \cite{GerdtKhvedelidzePalii2009}.

\subsection{The global orbit space $\mathfrak{P}(\mathbb{R}^{8}) / \mathrm{SU}(3)\,$}

\begin{figure}
\begin{center}
    \includegraphics[scale=0.7]{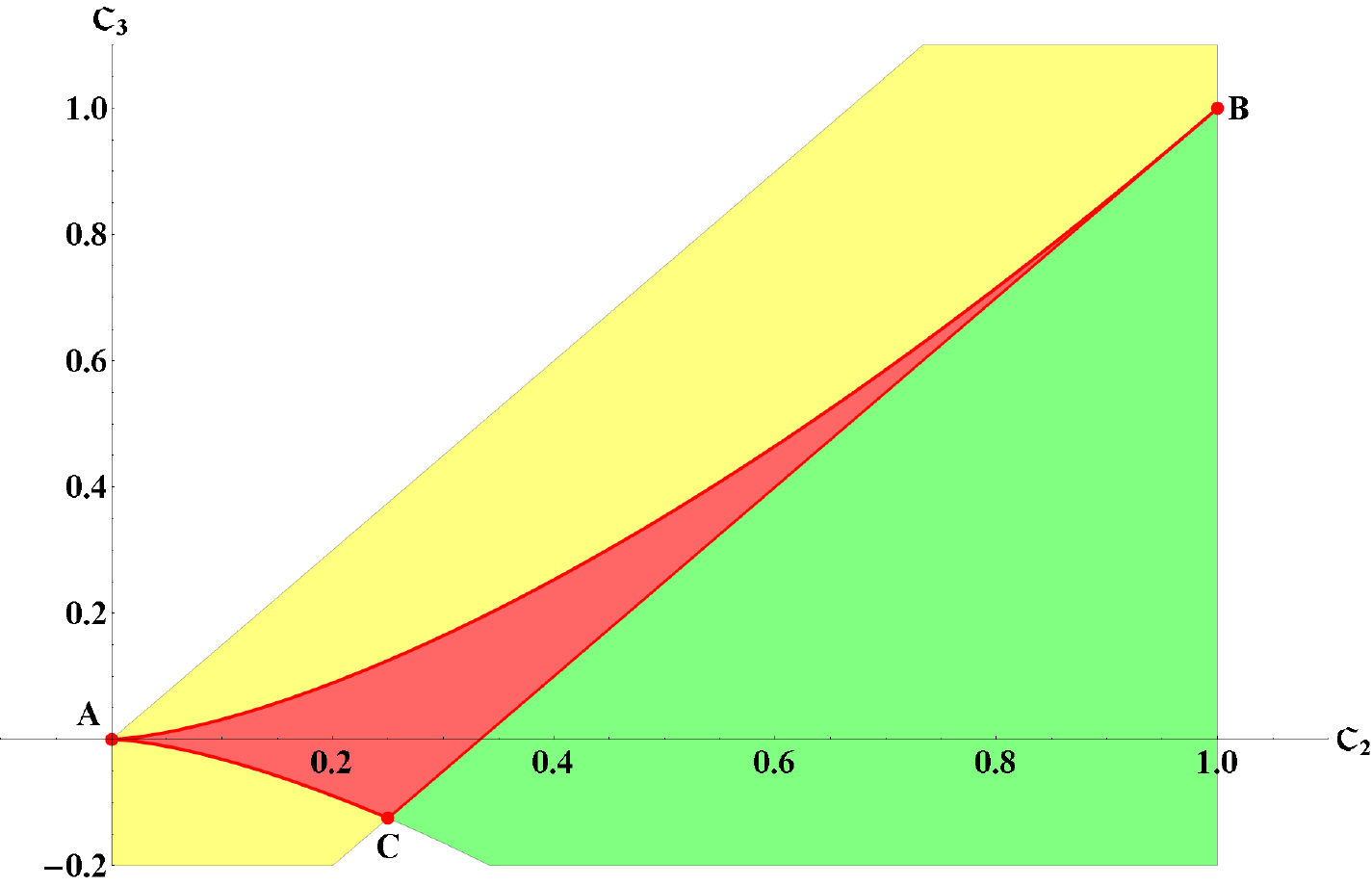}
    \caption{Triangle ABC as qutrit's global orbit space on the Casimir's
    $(\mathfrak{C}_2\,,\mathfrak{C}_3)$\--plane.}\label{Fig:QuritSpace}
\end{center}
\end{figure}

\noindent$\bullet$ {\bf The semi-positivity of density matrix}$\bullet$ The equations
(\ref{eq:pos1}) and (\ref{eq:pos2}) defining the
semi-positivity of the qutrit density matrix in terms of the Bloch vector $\boldsymbol{\xi}$ can be rewritten via two SU(3) Casimir invariants  $ \mathfrak{C}_2$
and $  \mathfrak{C}_3$ as follows
\begin{eqnarray}\label{eq:ineqCasimir1}
&&0 \leq \mathfrak{C}_2 \leq 1 \,,\\
&&0 \leq 3\mathfrak{C}_2-2\mathfrak{C}_3\leq 1\,.
\label{eq:ineqCasimir2}
\end{eqnarray}
$\bullet$ {\bf The Hermicity of density matrix}$\bullet$
The inequalities (\ref{eq:ineqCasimir1}) and (\ref{eq:ineqCasimir2})   should be completed by the reality condition of eigenvalues of the  qutrit density matrix expressed as polynomial inequality in two Casimirs. The latter  represents the non-negativity requirement for the discriminant  of the characteristic equation $\det\left(\lambda- \varrho\right) =0$  for the qutrit  density matrix $\varrho$:
\begin{equation}
\label{eq:DiscPos}
\mbox{Disc}:=
\mathfrak{C}_2^3-\mathfrak{C}_3^2 \geq  0\,.
\end{equation}
Thus the intersection of  the strip defined  by the linear inequalities (\ref{eq:ineqCasimir1}) and (\ref{eq:ineqCasimir2}) with the domain (\ref{eq:DiscPos})  determines  the
qutrit  state space $\mathfrak{P}(\mathbb{R}^{8}) \,.$ This intersection represents  the curvilinear  triangle ABC on the $(\mathfrak{C}_2, \mathfrak{C}_3)$\--plane
depicted on the Figure \ref{Fig:QuritSpace}.

Now, we show that triangle ABC is nothing else as the coset space $\mathfrak{P}(\mathbb{R}^{8}) / \mathrm{SU}(3)\,.$ for the qutrit state space. Indeed, since  the determinant of the Procesi-Schwarz $\mbox{Grad}_{\mathrm{SU}(3)}$\--matrix
\begin{equation}\label{eq:GradSU3}
\mbox{Grad}_{\mathrm{SU}(3)}=\left(
    \begin{array}{cc}
     4 \mathfrak{C}_2 &  6\mathfrak{C}_3 \\[2mm]
      6 \mathfrak{C}_3 & 9 \mathfrak{C}_2^2
    \end{array}
\right)
\end{equation}
is proportional to the discriminant (\ref{eq:DiscPos})
\[
\det||\mbox{Grad}_{\mathrm{SU}(3)}||= 36(\mathfrak{C}_2^3-\mathfrak{C}_3^2)\,,
\]
the semi-positivity  of Grad\--matrix, that determines the orbit space $\mathfrak{P}(\mathbb{R}^{8}) / \mathrm{SU}(3)\,$ coincides with the  Hermicity requirement
of the qutrit density matrix.

\subsection{The orbit  space $\mathfrak{P}/\mathrm{ SU(2)\times U(1)} $}
\label{subsec: qutri/u1t}

Let us start with the observation that the  $\mathrm{SU}(3)$ Casimir invariants can be expressed in terms of   the four  $\mathrm{ SU(2)\times U(1)}$\--invariants (\ref{eq:f1})-(\ref{eq:f4}) as
\begin{equation}\label{eq:CInvL}
 \mathfrak{C}_2=f_1^2+f_2+f_3\,, \qquad
 \mathfrak{C}_3=f_1(f_2-\frac{1}{2}f_3)-
 \frac{3\sqrt{3}}{4} f_4 - f_1^3\,.
\end{equation}
Because we are interested in the projection of orbit space $\mathfrak{P}/\mathrm{ SU(2)\times U(1)} $ to the space $\mathfrak{P}(\mathbb{R}^{8}) / \mathrm{SU}(3)\,,$ it is constructive to use relations (\ref{eq:CInvL})  and to build the integrity basis that contains
$ \mathfrak{C}_2$ and $ \mathfrak{C}_3$ as its elements of the second and third degree:
\[
{\cal{P}}^{\mathrm{ SU(2)\times U(1)} }:=\{f_1,\,f_2,\,\mathfrak{C}_2,\,\mathfrak{C}_3\}\,.
\]
As calculations show
the $4\times 4$ Grad\--matrix for  the integrity
basis  $\{f_1,\,f_2,\,\mathfrak{C}_2,\,\mathfrak{C}_3\}\,$ can be written  in the block form
\begin{equation}\label{eq:GradSU2u1}
\mbox{Grad}_{\mathrm{SU}(2)\times \mathrm{U}(1)}=\left(
    \begin{array}{cccc}
     { \cal A} ,&   { \cal B} \\[2mm]
       { \cal  B}^T, &   { \cal D} \\[2mm]
    \end{array}\right)\,,
\end{equation}
with $  { \cal A} :=\mathrm{diag}(1, 4f_2)\,,$ matrix
${ \cal D}$ denotes the SU(3) Grad\--matrix (\ref{eq:GradSU3}) and
\begin{equation}
  { \cal B}:=\left(
\begin{array}{cc}
2f_1, &  \tfrac{3}{2}(3 f_2 - f_1^2 - \mathfrak{C}_2)\\[2mm]
4f_2, & 3 f_1 ( f_2 +  \mathfrak{C}_2 ) + 2 \mathfrak{C}_3\\[2mm]
\end{array}
\right)\,.
\end{equation}
It is easy to see
that the semi-positivity of matrix (\ref{eq:GradSU2u1}) is reduces to the non-negativity condition for its determinant:
\begin{equation}
\det|| \mbox{Grad}_{\mathrm{SU}(2)\times \mathrm{U}(1)}|| \geq 0
\end{equation}
Furthermore, from the expression
\begin{equation}
\begin{split}
\det|| \mbox{Grad}_{\mathrm{SU}(2)\times \mathrm{U}(1)}||=&
4 \left(\mathfrak{C}_2+3 f_2-f_1^2\right)\times\\
&\times \left[-9 f_1^2 \left(\mathfrak{C}_2^2+3 f_2^2\right)-12 \mathfrak{C}_3 f_1 (\mathfrak{C}_2-3 f_2) \right. \\
& \qquad \left. +3 f_1^4 (2 \mathfrak{C}_2+3 f_2)+27 f_2 (\mathfrak{C}_2-f_2)^2-4 \mathfrak{C}_3^2+4 \mathfrak{C}_3
   f_1^3-f_1^6\right].
\end{split}
\end{equation}
it follows that domain of the Grad\--matrix non-negativity  is the 4-dimensional body bounded by two 3-dimensional hypersurfaces that we denote by $\Sigma_+$ and $\Sigma_-$.  The explicit parametrization  of  $\Sigma_\pm$  can be found  by solving  the equation
\begin{equation}
-9 f_1^2 \left(\mathfrak{C}_2^2+3 f_2^2\right)-12 \mathfrak{C}_3 f_1 (\mathfrak{C}_2-3 f_2-\frac{1}{3}f_1^2)
 +3 f_1^4 (2 \mathfrak{C}_2+3 f_2)+27 f_2 (\mathfrak{C}_2-f_2)^2-4 \mathfrak{C}_3^2-f_1^6=0
\end{equation}
with respect to $ \mathfrak{C}_3 $.
Thereby, the $\Sigma_\pm$ hypersurfaces  are  given by equations:
\begin{equation}\label{eq:Sigmpm}
\mathfrak{C}_3=
\frac{3}{2} \left(f_1 (3 f_2-\mathfrak{C}_2)+\frac{f_1^3}{3} \mp \sqrt{3 f_2} \left(-\mathfrak{C}_2+f_2+f_1^2\right)\right).
\end{equation}
According to (\ref{eq:Sigmpm}), the $\Sigma_+$ and $\Sigma_-$ intersect if
\begin{equation}
\sqrt{3 f_2} \left(f_2+f_1^2-\mathfrak{C}_2\right)=0.
\end{equation}
Thus,  $\Sigma_\pm$ hypersurfaces intersect
 along the following 2-dimensional surfaces $\Delta_1$ and  $\Delta_2$ :
\begin{enumerate}
\item $\Delta_1$ surface :
\begin{equation}
f_2=0\,, \qquad \mathfrak{C}_3=\frac{3}{2}f_1
\left( \frac{f_1^2}{3} - \mathfrak{C}_2 \right),
\end{equation}
\item $\Delta_2$ surface :
\begin{equation}
f_2+f_1^2-\mathfrak{C}_2=0\,, \qquad \mathfrak{C}_3=
3f_1\left( \mathfrak{C}_2 - \frac{4}{3}f_1^2\right)\,.
\end{equation}
\end{enumerate}

\begin{figure}
\begin{center}
	\includegraphics[scale=0.7]{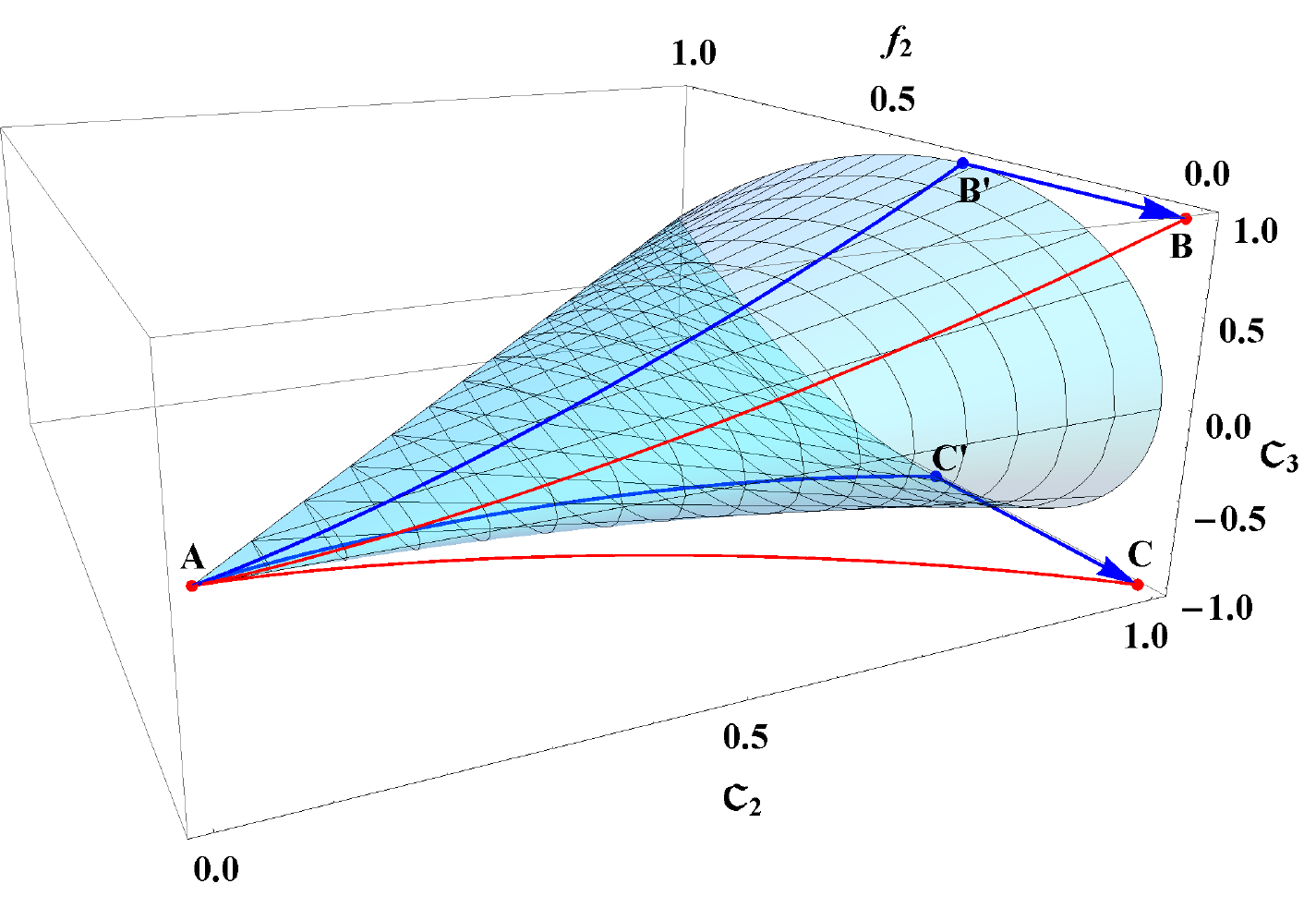}\\
    \caption{Domain
    $\mbox{Grad}_{\mathrm{SU}(2)\times \mathrm{U}(1)} \geq 0$
    and its projection to $(\mathfrak{C}_2,\;\mathfrak{C}_3)\,$ for
   $f_1=0$\,.}\label{fig:GradPos=0}
\end{center}
\end{figure}

To make description of orbit space  more transparent,  consider its  3-dimensional
cross sections  for different values of  the ``local'' invariant  $f_1$:

\noindent$\bullet$ {\bf  $\mathfrak{P}/\mathrm{ SU(2)\times U(1)} $ for  $f_1=0$ }$\bullet$
The 3-dimensional slice of   the  ``local'' orbit space fixed by the local invariant
$f_1=0$ is drawn on the Figure \ref{fig:GradPos=0}.
From this picture one can see that the projection of the ``cone of semipositivity'' of the
Grad\- matrix to the
$(\mathfrak{C}_2\,, \mathfrak{C}_3)\,$\--plane reproduces exactly the
ABC triangle,  the orbit space $\mathfrak{P}(\mathbb{R}^{8}) / \mathrm{SU}(3)$
depicted on Figure \ref{Fig:QuritSpace}.

For non-vanishing values of $f_1$ the  attainable area of the Casmir invariants $(\mathfrak{C}_2\,, \mathfrak{C}_3)\,$ is shrinking.
To illustrate this effect,  we give below the corresponding pictures for positive, $f_1=2/5$ and negative,  $f_1=-2/5$
values of invariant $f_1$.

\begin{figure}
\begin{center}
	\includegraphics[scale=0.7]{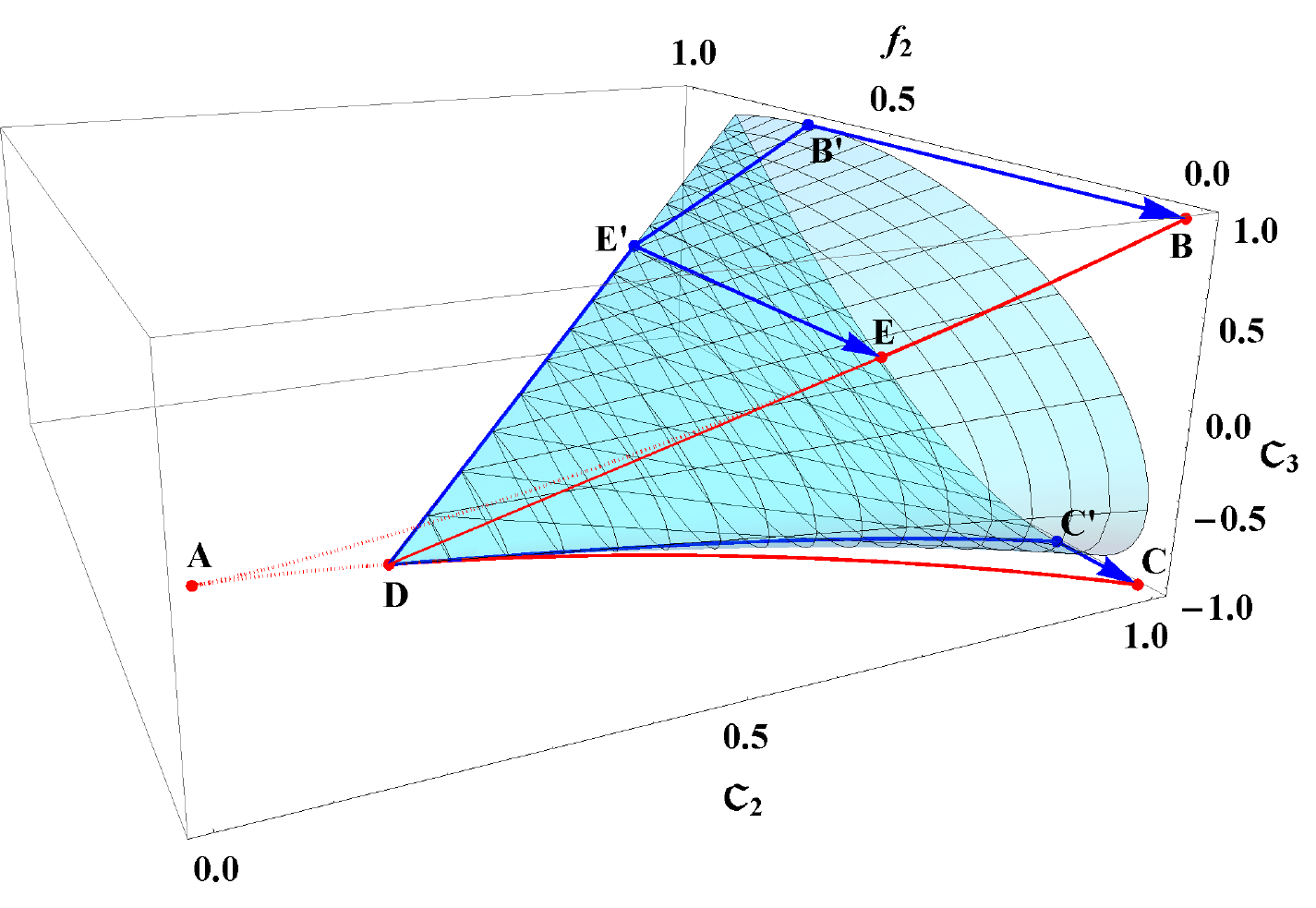}\\
    \caption{Domain
    $\mbox{Grad}_{\mathrm{SU}(2)\times \mathrm{U}(1)} \geq 0$
    and its projection to $(\mathfrak{C}_2,\;\mathfrak{C}_3)\,$ for
   $f_1=2/5$\,.}\label{fig:GradPos=2/5}
\end{center}
\end{figure}

\begin{figure}
\begin{center}
    \includegraphics[scale=0.7]{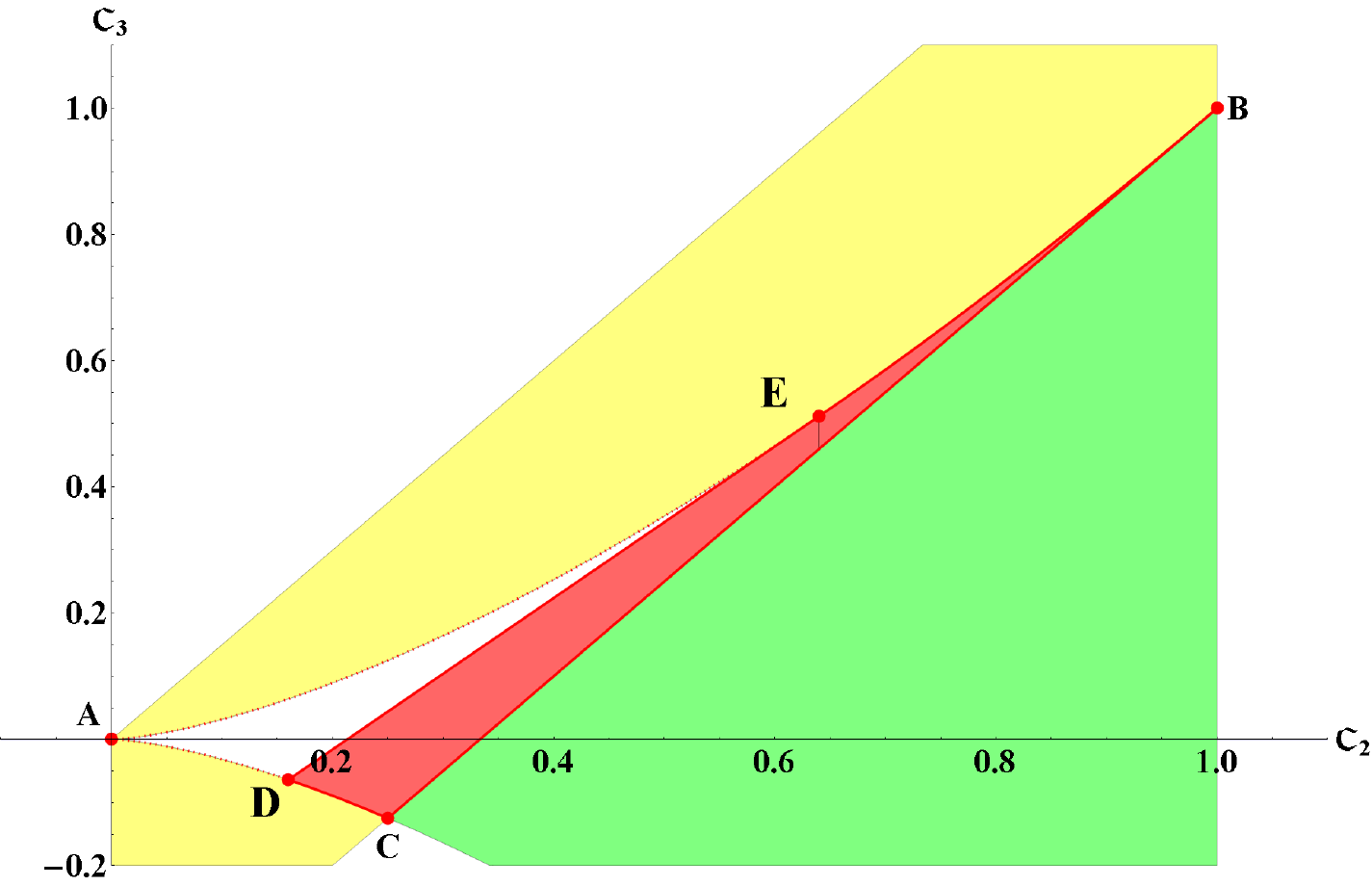}
    \caption{DCBE is the image of $\mathfrak{P}/\mathrm{ SU(2)\times U(1)} $ on SU(3) orbit space for fixed
     $f_1=2/5$\,.}\label{Fig:+025}
\end{center}
\end{figure}

\noindent$\bullet$ {\bf  $\mathfrak{P}/\mathrm{ SU(2)\times U(1)} $ for  $f_1=2/5$ }$\bullet$ For this value the ``cone of semipositivity'' is  drawn on the  Figure
\ref{fig:GradPos=2/5}. For non-zero  values of $f_1$ the vertex
 of ``cone of semipositivity'' intersects
the Casmir invariants $(\mathfrak{C}_2\,, \mathfrak{C}_3)$\-- plane point D that differ from point A.
The line DE  is projection of the surface  $\Delta_2$ with $f_1=2/5$.  With growing $f_1$
the line DE moves towards BC  and for $ f_1 =1/2  \,$ it covers the last.
To make illustration  the ``shrinking effect''
more vivid, the allowed domaib for SU(3) Casimirs invariants is shown on  the Figure \ref{Fig:+025}.

\begin{figure}
\begin{center}
	\includegraphics[scale=0.7]{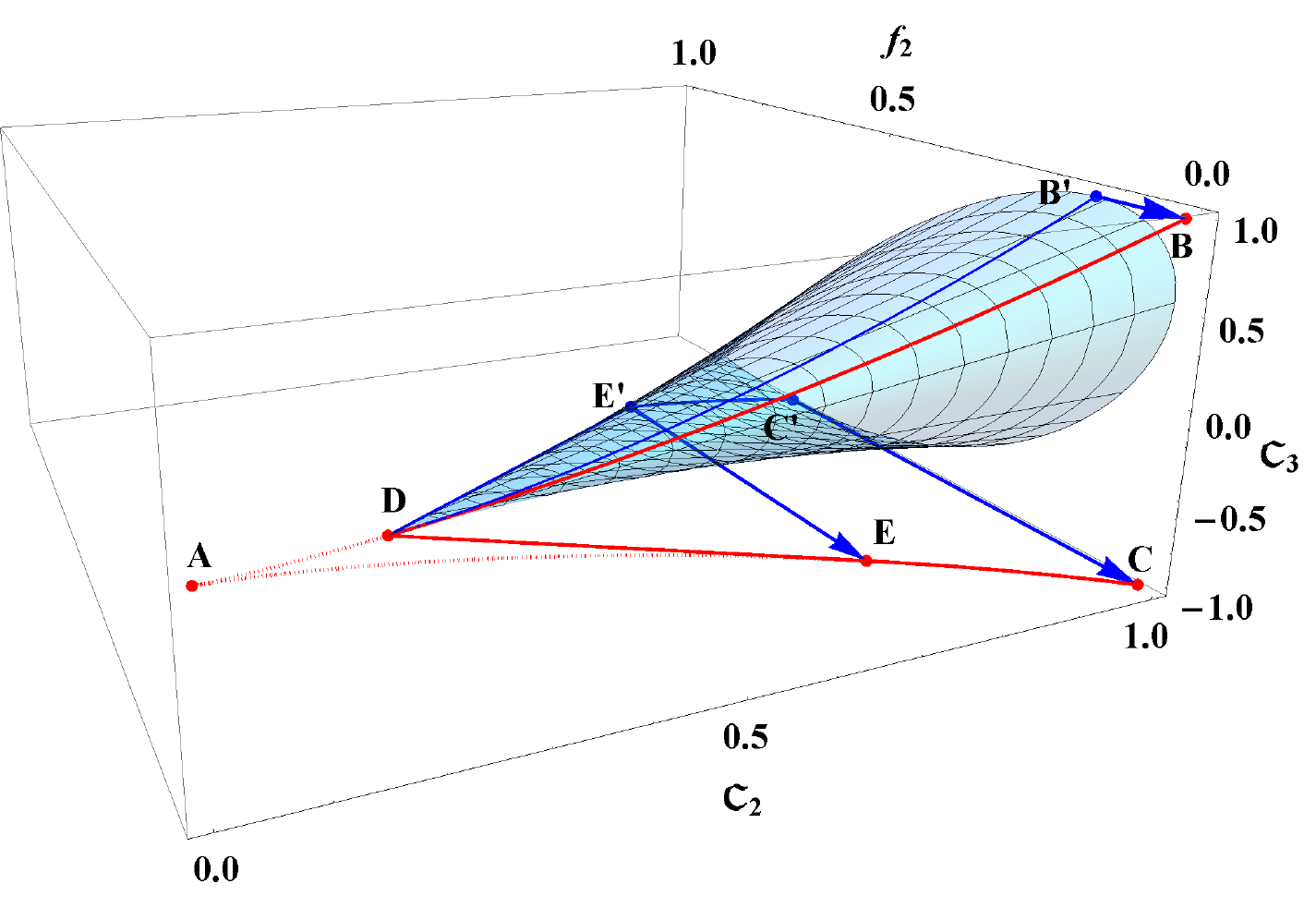}\\
    \caption{Domain
    $\mbox{Grad}_{\mathrm{SU}(2)\times \mathrm{U}(1)} \geq 0$
    and its projection to $(\mathfrak{C}_2,\;\mathfrak{C}_3)\,$ for
   $f_1=-2/5$\,.}\label{fig:GradPos=-2/5}
\end{center}
\end{figure}

When the ``local'' invariant $f_1 $ lies in the interval $(0, -1], $ an  alternative  mechanism of
shrinking of the triangle ABC triangle is realized:

\noindent$\bullet$ {\bf  $\mathfrak{P}/\mathrm{ SU(2)\times U(1)} $ for  $f_1=-2/5$ }$\bullet$ For this case the
the ``cone of semi-positivity'' is  depicted  on the  Figure
\ref{fig:GradPos=-2/5}. When $ f_1 $ takes negative values the points D and E move toward the point B and all coincide  for $f_1=-1\,.$
\begin{figure}
\begin{center}
    \includegraphics[scale=0.7]{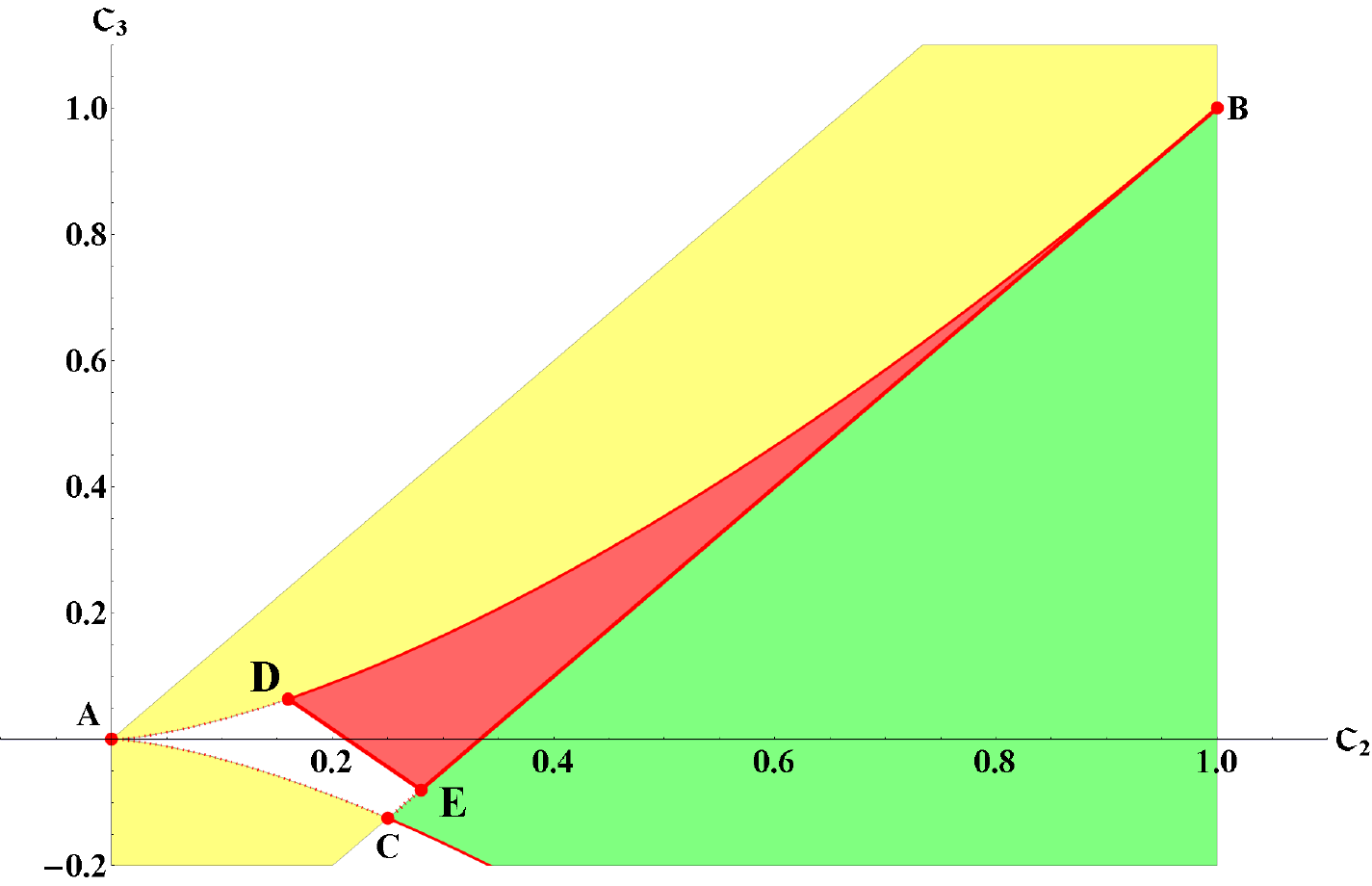}
    \caption{DBE is the  image of $\mathfrak{P}/\mathrm{ SU(2)\times U(1)} $ on SU(3) orbit space for
     $f_1=- 2/5$\,.}\label{Fig:-025}
\end{center}
\end{figure}
The  Figure \ref{Fig:-025} exemplifies the effect of shrinking of  the allowed SU(3) Casimirs invariants domain for negative value $f=-2/5.$

\begin{figure}
\begin{center}
    \includegraphics[scale=0.69]{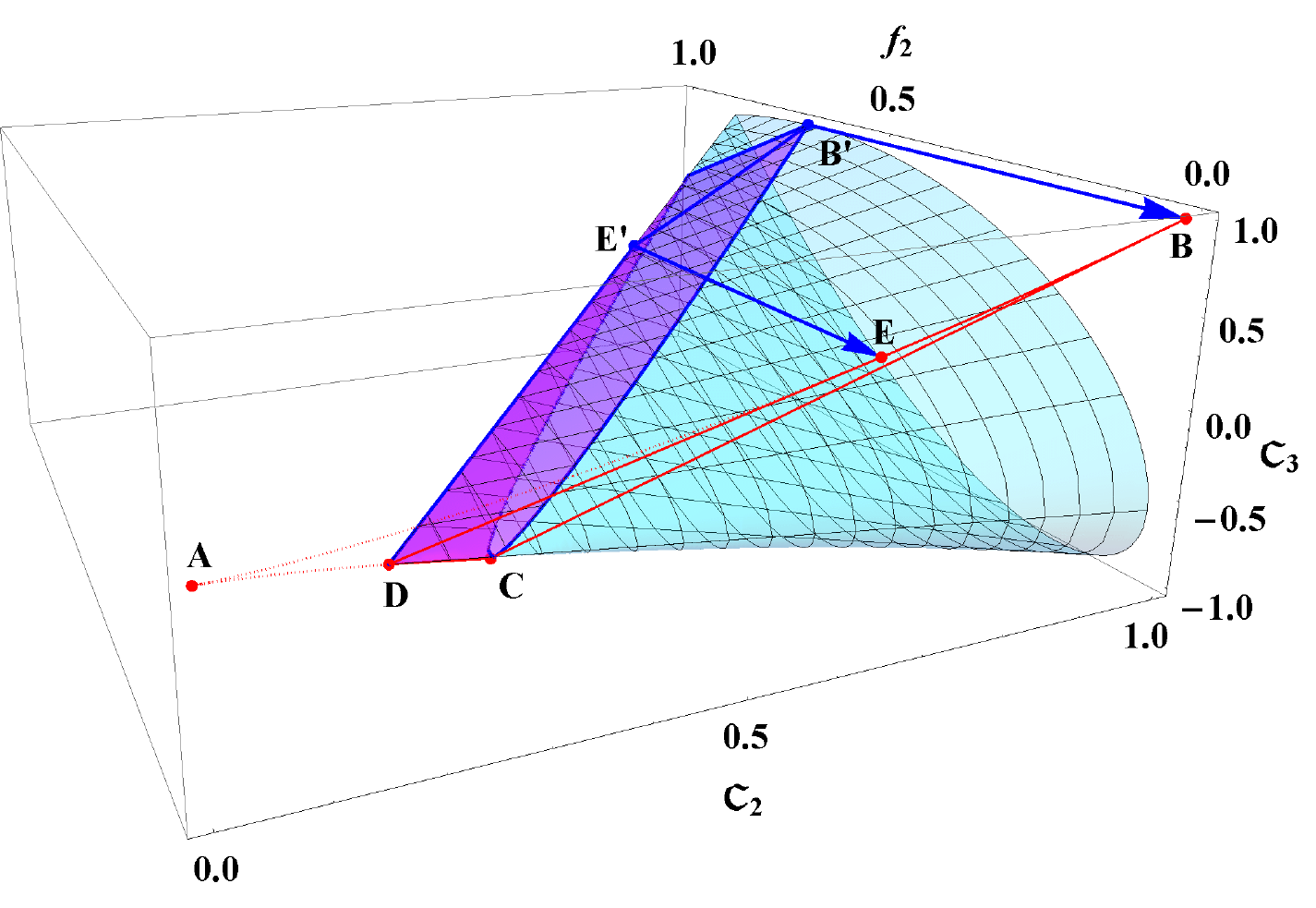}
    \includegraphics[scale=0.69]{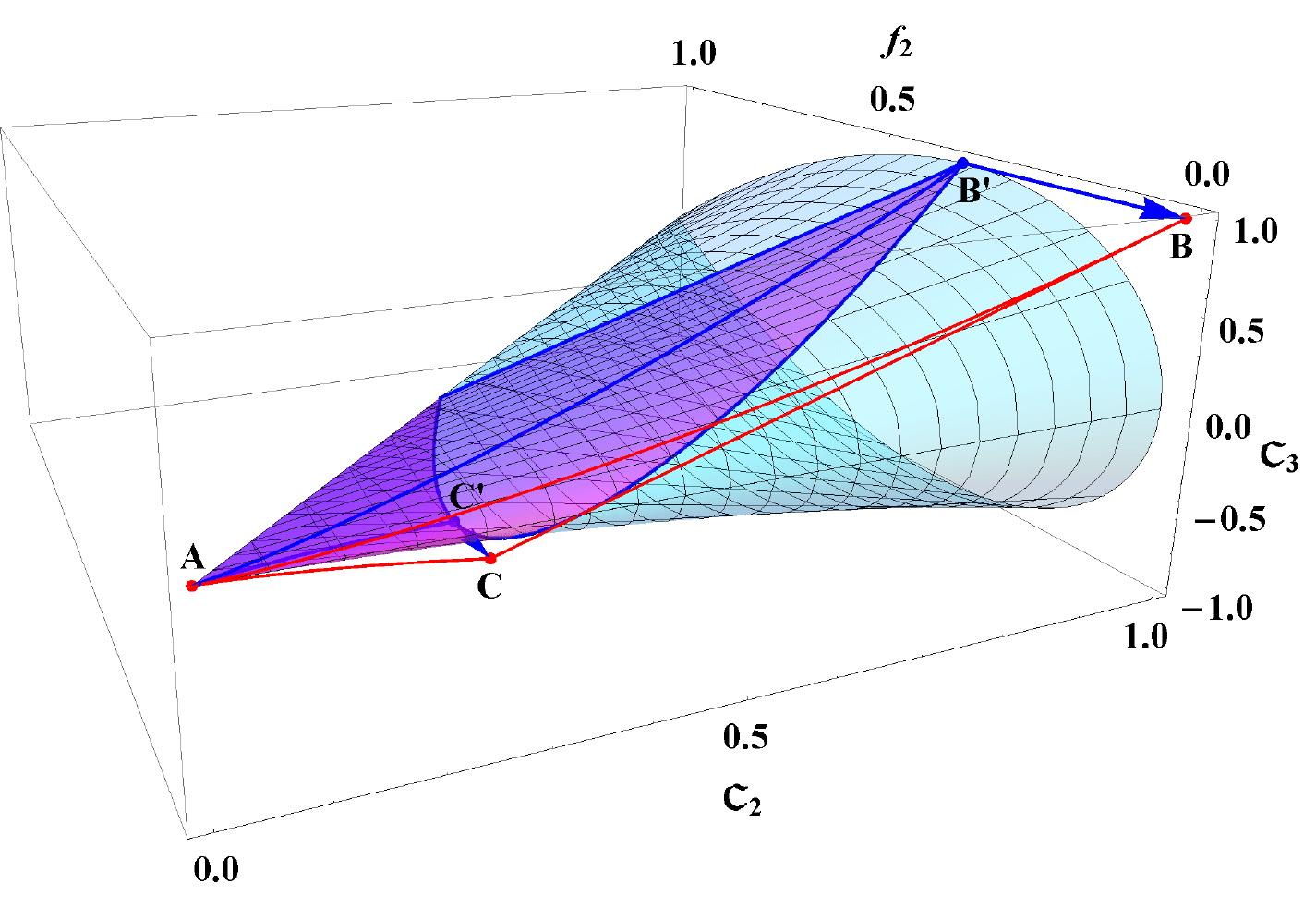}\\
    \includegraphics[scale=0.69]{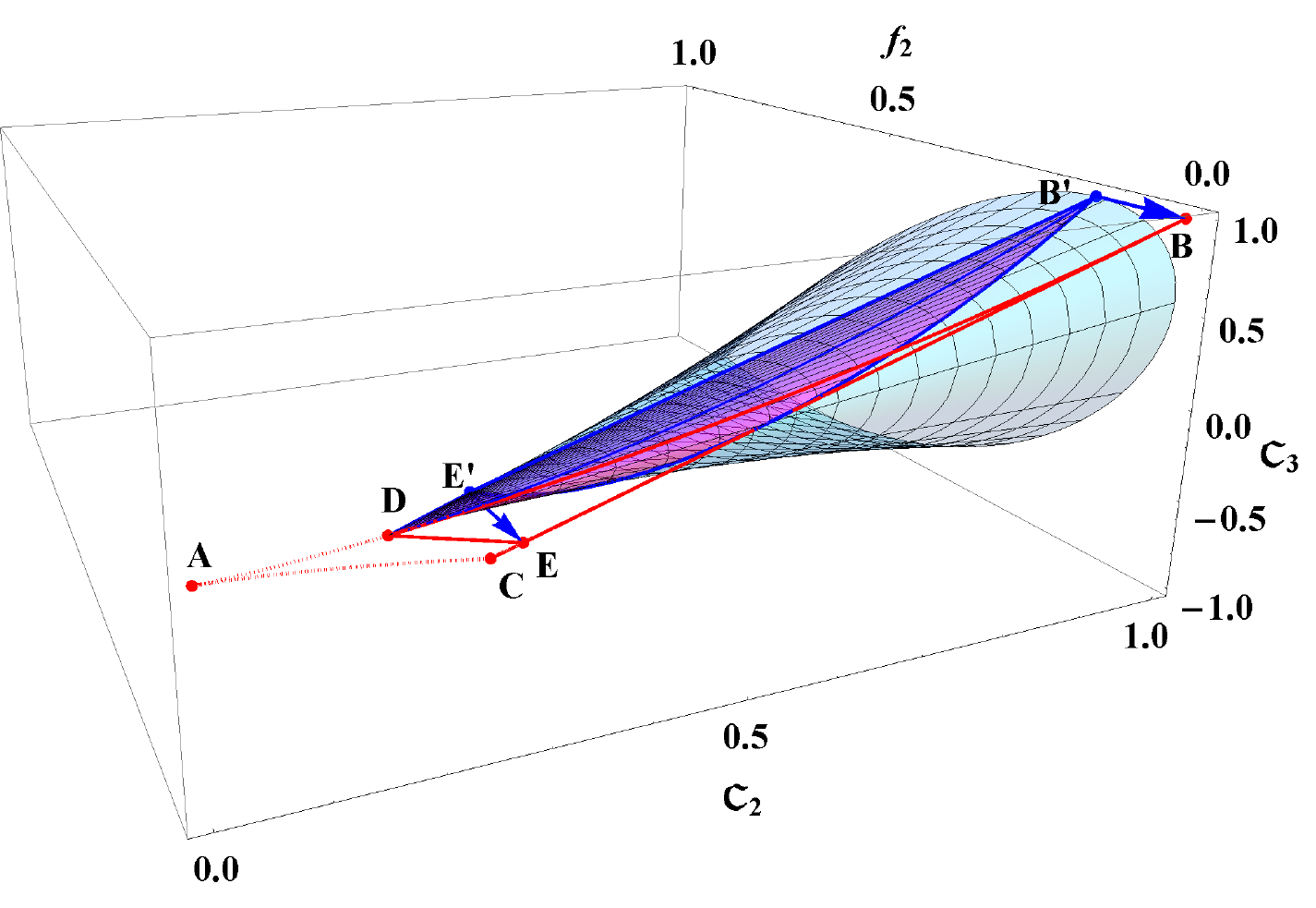}
    \caption{$\mathfrak{P}/\mathrm{ SU(2)\times U(1)} $
     slices for  $f_1=2/5\,$ (top),  $f_1=0\,$ and $f_1=-2/5\,$ (bottom).}\label{Fig:Orbsu2ui}
\end{center}
\end{figure}

Finally, the  3-dimensional slices of the orbit space $\mathfrak{P}/\mathrm{ SU(2)\times U(1)} $ for different values of $f_1$ are presented on  the Figure  \ref{Fig:Orbsu2ui}.

\section{Conclusion}

In the present note we analyze the $\mathrm{SU}(2)\otimes\mathrm{U}(1)$\--orbit space
of qutrit  treating it as simplified  analogue  of  the entanglement space
of a composite system. The qutrit orbit space
is described as a semialgebraic variety in  $\mathbb{R}^4\,,$
defined by a set of polynomial inequalities in $\mathrm{SU}(2)\otimes\mathrm{U}(1)$
adjoint invariants.
These inequalities follow from the simultaneous semi-positivity of two  matrices,
the  qutrit density matrix and the Procesi-Schwarz Grad\- matrix, constructed  with the aid of fundamental set of   $\mathrm{SU}(2)\otimes\mathrm{U}(1)$\-- invariants.
It was discussed  in details how  the semi-positivity of the
$\mbox{Grad}\,$-matrix for $\mathrm{SU}(2)\otimes\mathrm{U}(1)$
invariants provides new restrictions  on the geometry of orbit
space  in contrast to the case of the $\mathrm{SU}(3)$ orbit
space.

\bigskip

\noindent $\bullet\,${\bf Acknowledgements} $\bullet\,$ The work is supported in part by the Ministry of Education
and Science of the Russian Federation (grant 3003.2014.2) and the Russian Foundation for Basic Research
(grant 13-01-00068).


\end{document}